\documentclass[12pt, a4paper]{article}
\usepackage[utf8]{inputenc}
\usepackage[T1]{fontenc}
\usepackage{graphicx}
\usepackage{authblk}
\usepackage{amsmath}
\usepackage{xcolor}
\usepackage{setspace}
\usepackage[round, semicolon]{natbib}
\setcitestyle{aysep={}}

\usepackage{lipsum}
\usepackage{cancel}
\usepackage{tabularx}
\usepackage{array}
\usepackage{pdflscape}
\usepackage{booktabs}

\title{Generalising the Central Dogma as a cross-hierarchical principle of biology}

\author[1,2,*]{Nobuto Takeuchi}
\author[3,2]{Kaneko Kunihiko}
\affil[1]{School of Biological Sciences, University of Auckland, Auckland, New Zealand}
\affil[2]{Research Center for Complex Systems Biology, Universal Biology Institute, University of Tokyo, Tokyo, Japan}
\affil[3]{Niels Bohr Institute, University of Copenhagen, Copenhagen, Denmark}
\affil[*]{Corresponding: nobuto.takeuchi@auckland.ac.nz}

\date{}

\begin{document}
\maketitle

\begin{abstract}
    The Central Dogma of molecular biology, as originally proposed by Crick,
    asserts that information passed into protein cannot flow back out. This
    principle has been interpreted as underpinning modern understandings of
    heredity and evolution, implying the unidirectionality of information flow
    from nucleic acids to proteins. Here, we propose a generalisation of the
    Central Dogma as a division of labour between the transmission and
    expression of information: the transmitter (nucleic acids)
    perpetuates information across generations, whereas the expressor (protein)
    enacts this information to facilitate the transmitter's function without
    itself perpetuating information. We argue that this generalisation offers
    two benefits. First, it provides a unifying perspective for comparing the
    Central Dogma to analogous divisions of labour observed at vastly different
    biological scales, including multicellular organisms, eukaryotic cells,
    organelles, and bacteria. Second, it offers a theoretical framework to
    explain the Central Dogma as an outcome of evolution. Specifically, we
    review a mathematical model suggesting that the Central Dogma originates
    through spontaneous symmetry breaking driven by evolutionary conflicts
    between different levels of selection. By reframing the Central Dogma as an
    informational relationship between components of a system, this
    generalisation underscores its broader relevance across the biological
    hierarchy and sheds light on its evolutionary origin.
\end{abstract}

The Central Dogma of molecular biology is one of the most widely known yet
frequently misunderstood concepts in biology. Many think it means information
flows from DNA to RNA to protein---since this is how it is taught in textbooks
\citep{Cox2013,Barton2007,Alberts2015,Phillips2013}. However, Crick's original
definition is, ``once `information' is passed into protein it cannot get out
again''   \citep{Crick1958}. This means, ``the transfer of information from
nucleic acid to nucleic acid, or from nucleic acid to protein may be possible,
but transfer from protein to protein, or from protein to nucleic acid is
impossible,'' where information means ``the precise determination of sequence,
either of bases in the nucleic acid or of amino acid residues in the protein''
\citep{Crick1958}. Thus, while the Central Dogma rules out the reading of
information from protein, it does not preclude reverse transcription, a process
that is often misconceived as reversing the Central Dogma
\citep{Reversed1970,Crick1970}.

The Central Dogma has been interpreted as underpinning the modern understanding
of biological evolution, in particular, by providing the molecular basis for the
non-inheritance of acquired characters. For example, Monod highlighted the
Central Dogma's role in establishing the independence of the genetic message
from external influences, stating \citep{Judson1996a}:
\begin{quote}
    Since the whole Darwinian concept is based on change through the inheritance
    of new traits, on selection pressing on a somewhat varied population, so
    long as you could not say exactly how inheritance occurred, physically, and
    what the generator of variety was, chemically, Darwinism was still up in the
    air. So, what molecular biology has done, you see, is to prove beyond any
    doubt but in a totally new way the complete independence of the genetic
    information from events occurring outside or even inside the cell---to prove
    by the very structure of the genetic code and the way it is transcribed that
    no information from outside, of any kind, can ever penetrate the inheritable
    genetic message. This was believed but never proved until the structure of
    DNA and the mechanisms of protein synthesis were understood. This was what
    Francis called the Central Dogma: no information goes from protein to DNA.
\end{quote}
Maynard Smith also wrote \citep{MaynardSmith2000}:
\begin{quote}
    I think that the non-inheritance of acquired characters is a contingent
    fact, usually but not always true, not a logical necessity. Insofar as it is
    true, it follows from the ``central dogma'' of molecular biology, which
    asserts that information travels from nucleic acids to proteins, but not
    from proteins to nucleic acids.
\end{quote}

These quotes display a subtle shift in the conceptualization of the Central
Dogma in the context of evolution. Crick defined the Central Dogma as the
impossibility of retrieving information from protein, focusing on a property of a single type of molecule. In contrast, Monod and Maynard Smith
reconceptualised the Central Dogma as the irreversibility of information flow from nucleic
acids to proteins, focusing on a relationship between different types of molecules. This shift of focus from the property of a single molecule to the relationship between multiple molecules can be
furthered to formulate a beneficial abstraction of the Central Dogma, as
discussed below.

\section{The Central Dogma as a division of labour between information transmission and expression}

We propose to abstract the Central Dogma as a division of labour between the
transmission and expression of information. The transmitter perpetuates
information by maintaining persistent lines of descent, whereas the expressor
enacts this information to facilitate the transmitter's function without
directly perpetuating it. This division of labour abstracts the Central Dogma,
with nucleic acids as the transmitter and protein as the expressor. This division implies
that the flow of information from nucleic acids to protein is irreversible---if it were
reversible, protein would extend its lines of descent interwoven with those of
nucleic acids, perpetuating information. However, this division does not exclude the
reading of information from protein, generalising the scope of the Central
Dogma. For example, the replication of protein sequences (i.e., protein-to-protein information
transfer) is permissible as long as it remains transient and does not produce
persistent lines of descent.\footnote{The protein replication discussed in the main text excludes prions. However, prions provide an illuminating thought exercise in relation to our argument. Unlike nucleic acids, prions transmit structural rather than sequence information, yet they can still display the division of labour between information transmission and expression. Suppose a protein has three distinct folds: a natural fold, a template misfold, and a catalytic misfold. The template misfold transmits structural information by serving as a ``template'' to convert the natural fold into additional copies of itself. The catalytic misfold catalyses this structural conversion. The catalytic misfold is generated through the irreversible structural transformation of the template misfold.} In addition to generalising the Central Dogma, this
division also introduces an additional attribute that is not mentioned in the
Central Dogma: protein enacts information in nucleic acids to help them perpetuate this
information.

The division of labour between information transmission and expression
represents a specialised form of the reproductive division of labour. The
reproductive division of labour means that some individuals within a social
group specialise in reproduction, whereas the others perform cooperative tasks
such as foraging or defence \citep{Keller2014}. However, individuals performing
cooperative tasks may sometimes become reproducers, for example, when
pre-existing reproducers die \citep{Cant2014}. By contrast, the
transmitter-expressor division of labour imposes a constraint that the expressor
does not become the transmitter, ensuring the unidirectionality of information
transfer from the transmitter to the expressor.

Abstracting the Central Dogma as the division of labour between information
transmission and expression offers two benefits. First, it provides a unifying
perspective that connects the Central Dogma to analogous divisions of labour
observed across vastly different biological scales. Second, it offers a
theoretical framework to explain the Central Dogma as a logical consequence of
evolution, rather than a contingent historical fact, as suggested by Maynard
Smith. These advantages are explored in the following sections.

\section{The Central Dogma as a particular instance of the cross-hierarchical principle}

The Central Dogma is one example of the division of labour between information
transmission and expression. This division has evolved at nearly every
conceivable level of biological individuality.
Below, we explore how this division manifests across biological hierarchies, as summarised in Table~\ref{tbl:universality} (which provides a concise overview, allowing readers to proceed directly to the next section if they wish).

\begin{landscape}
\renewcommand{\arraystretch}{1.4} 
\newcolumntype{P}[1]{>{\raggedright\arraybackslash}p{#1}}
\begin{table}[tbp]
    \centering
    \begin{tabularx}{\linewidth}{P{0.21\linewidth}P{0.12\linewidth}P{0.11\linewidth}X}
        \toprule
        \textbf{Biological scale} (elements in a group) & \textbf{Information transmitter} & \textbf{Information expressor} & \textbf{Division of labour between information transmission and expression}
        \\\midrule
        Multicellular organisms in a eusocial colony  & Queens               & Workers & Queens lay eggs. Workers perform tasks, such as foraging and brood care to enhance queens' reproductive success. Workers are sterile.
        \\
        Eukaryotic cells in a multicellular organism & Germline cells       & Somatic cells & Germline cells produce gametes. Somatic cells specialise in functions supporting organismal reproduction. Somatic cell lineages do not produce gametes.
        \\
        Organelles (nuclei) in a ciliate cell               & Micronuclei          & Macronuclei & Micronulei are capable of meiosis. Macronuclei are transcriptionally active. Macronuclei derive from micronuclei via genome fragmentations and truncations. 
        \\
        Bacterial cells in a colony of \textit{S.\,coelicolor}           & Wild-type cells            & Mutant cells & The wild type replicates substantially faster than the mutants. Mutants produce antibiotics. Mutants are produced via large-scale genome truncations.
        \\
        Molecules in a cell                & Nucleic acids                  & Protein & The Central Dogma of molecular biology. Nucleic acids transmit information across generations. Proteins provide catalysis. No reverse translation occurs.
        \\
        \bottomrule
    \end{tabularx}
    \caption{\label{tbl:universality}The division of labour between the transmission and expression of information occurs at distinct biological scales, illustrating the universality of this pattern across the biological hierarchy. The Central Dogma of molecular biology is an example of this division at the molecular level (bottom row). \textit{S.\ coelicolor}, \textit{Streptomyces coelicolor}.}
\end{table}
\end{landscape}

At the level of multicellular organisms, the division of labour between
information transmission and expression is exemplified by the differentiation
between queens and workers in eusocial animals, such as bees and ants
\citep{Hoelldobler1990,Bourke2011,Keller2014}. Queens lay eggs that develop into
workers or next-generation queens, perpetuating genetic information. By
contrast, workers are sterile and focus on tasks supporting the queens'
reproduction, such as foraging for food, defending the nest, and caring for the
brood. Therefore, queens are the transmitter, while workers are the expressor.

Another example of this division of labour at the level of multicellular
organisms is provided by siphonophores, a class of colonial marine invertebrates
related to corals and jellyfish \citep{Bourke2011, Mapstone2014}. These animals
display a differentiation between reproductive and non-reproductive `zooids,'
which are the clonal units that compose a colony and are homologous to the
individuals of solitary jellyfish. In many siphonophores (specifically in the
Codonophora group), a batch of zooids clonally develops from each bud that is
sequentially generated in the colony's growth zone \citep{Dunn2005}. Each batch
of zooids then differentiates into a defined organisation of reproductive and
non-reproductive zooids. Reproductive zooids produce gametes, while
non-reproductive zooids are specialised for feeding, defence, etc. Therefore,
the growth zone, along with its descendant reproductive zooids, is the
transmitter, while the non-reproductive zooids are the expressor. However, to
our knowledge, it is unknown whether the differentiation between reproductive
and non-reproductive zooids is irreversible.

At the eukaryotic cell level, the division of labour between information
transmission and expression is exemplified by the differentiation between
germline and somatic cells in multicellular organisms
\citep{Buss1987,Kirk1998,Bourke2011}. Germline cells produce gametes,
perpetuating genetic information, whereas somatic cells perform functions that
support the germline's dissemination of gametes. While some somatic cells, such
as somatic stem cells, are capable of replication (i.e., soma-to-soma
information transmission), they do not differentiate into germline cells and
thus do not produce gametes. This irreversible differentiation ensures the
unidirectionality of information transfer. Therefore, germline cells are the
transmitter, while somatic cells are the expressor.

We add that germline-soma distinction should be distinguished from early
germline segregation, which is a subset of germline-soma distinction
\citep{Solana2013,Devlin2023}. Early germline segregation refers to the
separation of germline cell lineages from somatic cell lineages at early stages
of development, thereby preventing germline cells from contributing to the production of
somatic cells during most of the organism's lifetime. In contrast, germline-soma
distinction does not depend on when---or even whether---germline cell lineages
are segregated from somatic cell lineages. For example, in sponges, which are a
basal group of metazoans, the stem-cell system consisting of archaeocytes and
choanocytes is capable of producing all cell types, including somatic cells and
gametes, throughout sponges' lifetime \citep{Funayama2013,Funayama2018}. In
contrast, somatic cells cannot differentiate into these stem cells or produce
gametes. Therefore, sponges display germline-soma distinction but lack germline
segregation \citep{Devlin2023}. In fact, early germline segregation occurs only
in a few taxa, such as vertebrates and insects, whereas germline-soma
distinction is widespread across metazoans \citep{Solana2013,Devlin2023}.

At the organellar level, the transmitter-expressor division of labour is
exemplified by the differentiation between micronuclei and macronuclei in
ciliates, a group of unicellular predators including \textit{Tetrahymena} and
\textit{Paramecium} \citep{Prescott1994,Ruehle2016,Cheng2020}. The micronucleus
contains two copies of each chromosome (diploid). In contrast, the macronucleus
contains numerous copies of subchromosomal DNA molecules, which are derived from
the chromosomes of the micronucleus through genome-wide site-specific
recombination, fragmentation, and truncation---processes that are likely to be
irreversible \citep{Prescott1994}. The micronucleus is transcriptionally silent,
whereas the macronucleus is active \citep{Prescott1994}. High DNA copy numbers
in the macronucleus are considered to enhance transcriptional throughput,
supporting large cell sizes, which are likely to be advantageous for predatory
lifestyles \citep{Cheng2020}. Ciliates reproduce both sexually and asexually.
During asexual reproduction, the micronucleus replicates through mitosis, while
the macronucleus replicates through `amitosis,' a process that stochastically
distributes DNA molecules to daughter cells after DNA replication
\citep{Prescott1994}. During sexual reproduction, the micronucleus replicates
through meiosis, and one of the resulting micronuclei is exchanged with another
cell via conjugation \citep{Ruehle2016}. Micronuclei from the two cells then
fuse, forming the micronucleus of the next generation. Meanwhile, the
macronucleus is completely destroyed and de novo reconstituted through the
replication and genome-wide restructuring of the new micronucleus
\citep{Ruehle2016}. This process ensures that the macronucleus is unable to
transmit genetic information through sexual reproduction. Therefore, the
micronucleus is the transmitter, while the macronucleus is the expressor.

At the bacterial cell level, the transmitter-expressor labour division is
exemplified by the differentiation between the wild-type and mutant
subpopulations in a colony of \textit{Streptomyces coelicolor}
\citep{Zhang2020}. The mutant subpopulations are derived from the wild type
through large-scale genome truncation, which deletes approximately $10^3$
genes---a process that is likely to be irreversible. The mutant subpopulations
display enhanced antibiotic production but reduced spore production compared to
the wild type. Therefore, the wild type is the transmitter, while the
genome-truncated mutants are the expressor. Another example at this level is the
`germ-soma' distinction in bacterial symbionts of multiple insect species
\citep{Buchner1965,Frank1996,Bressan2013}.

The examples outlined above illustrate that the division of labour between
information transmission and expression is an organisational principle observed
across many distinct biological scales, with the Central Dogma as an
example at the molecular level. Many of these examples are associated with a ``fraternal'' major evolutionary transition, a process in which individuals stay together after reproduction to form a higher-level unit of reproduction \citep{MaynardSmith1995,Queller1997,Bourke2011,Tarnita2013}. For example, germline-soma division occurs in multicellular organisms, where cells remain together after replication, and queen-worker division occurs in eusocial colonies, where individuals stay in their natal nests after maturation. The Central Dogma can be conceived as an evolutionary analogue of these divisions, corresponding to the major transition in which replicating molecules are grouped into collectives, such as protocells \citep{Takeuchi2019}. This perspective raises the question of
whether the origin of the Central Dogma can be explained by a general
evolutionary mechanism related to major transitions, a question we discuss in the next section.

\section{The Central Dogma as a consequence of multilevel selection}
Why does the Central Dogma hold? Crick states that it is highly improbable, for
stereochemical reasons, that protein-to-protein information transfer can occur
with the same simplicity as DNA-to-DNA transfer \citep{Crick1970}. Furthermore,
Crick notes that protein-to-RNA (or DNA) information transfer would require
complicated machinery that is entirely distinct from the one used for
RNA-to-protein information transfer and that there is no reason to believe that
such machinery might be needed by the cell \citep{Crick1970}.

Koonin suggests that protein folding is the fundamental cause of the Central
Dogma \citep{Koonin2015}. Koonin argues that protein folding prevents the
accurate retrieval of sequence information because amino acid residues that are
distant in the primary structure can become proximal in the tertiary structure.
Thus, retrieving information from protein requires denaturation. However,
denaturation generates misfolded globules, which are toxic for the cell and thus
need to be quickly degraded. Consequently, protein folding irreversibly blocks
the reading of amino-acid sequence information, causing the Central Dogma
\citep{Koonin2015}.

The hypotheses of Crick and Koonin described above are rooted in the
conceptualisation of the Central Dogma as a chemical property of single
molecules: once information is passed into protein, it cannot get out again.
Reconceptualising the Central Doma as a division of labour between the
transmission and expression of information offers an alternative framework to
explain its origin \citep{Takeuchi2019}. This framework describes how evolution
drives the differentiation of molecules into information transmitters (``genes'')
and expressors (``enzymes''), starting with undifferentiated molecules serving
both functions. This framework aligns with the RNA World hypothesis, which
posits that RNA fulfilled the dual roles of genes and enzymes before DNA and
protein evolved. However, while the RNA World hypothesis emphasises the unique
chemistry of RNA, this framework focuses on informational relationships
between molecules and is thus agnostic to their specific chemical properties.

\subsection{Mathematical model}
To explore the evolution of gene-enzyme differentiation, we review the
mathematical model introduced by \citet{Takeuchi2019}. The model assumes
molecules that are capable of functioning as both templates for replication and
catalysts for template replication, as is also assumed in the RNA world
hypothesis. Initially, all molecules are assumed to function as templates and
catalysts equally well. The model explores whether and how evolution leads to
gene-enzyme differentiation, which is defined by the emergence of the following
two asymmetries:
\begin{itemize}
    \item Catalytic asymmetry: enzymatic molecules provide catalysis, whereas
          genic molecules do not.
    \item Transmission asymmetry: genic molecules transmit information across
          generations, whereas enzymatic molecules do not.
\end{itemize}
To investigate this question, the model incorporates the following assumptions,
the rationale for which will be explained later:
\begin{enumerate}
    \item Providing catalysis is an ``altruistic'' trait.
    \item The population of molecules is ``stage-structured.''
    \item ``Multilevel selection'' operates.
\end{enumerate}

The first assumption, that providing catalysis is altruistic, is due to the
following trade-off: providing catalysis requires a molecule to fold into
specific secondary and tertiary structures, whereas serving as a template
requires these structures to unfold \citep{Durand2010,Ivica2013}. Therefore,
providing catalysis promotes the replication of other molecules at the expense
of the replication of self, thus representing an altruistic trait
\citep{Michod1983,Frank1994,Takeuchi2007a,Levin2017}.

The second assumption, that the population is stage-structured, posits that
molecules carrying equivalent information can exist in different stages (or
states) and express this information differently depending on the stage. For
example, within a population of molecules carrying equivalent information, those
in one stage might provide catalysis, while those in the other stage might not.
This stage-dependent catalysis abstracts the relationship between genes and
enzymes: they carry equivalent information related through the genetic code, but
only enzymes provide catalysis. Thus, stage-dependent catalysis allows the
evolution of catalytic asymmetry, one of the two asymmetries that characterise
gene-enzyme distinction (the other, transmission asymmetry, will be described
later).

\begin{figure}[tb]
    \centering
    \includegraphics{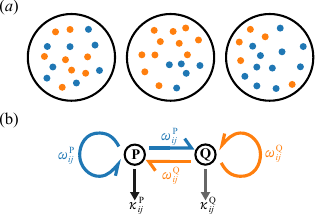}
    \caption{\label{fig:model}The model of gene-enzyme differentiation. (a)
    Replicating molecules (dots) are compartmentalised into protocells
    (circles), allowing multilevel selection to operate. (b) Molecule $j$ in
    stage $c\in\{\text{P}, \text{Q}\}$ in protocell $i$ has catalytic activity
    $\kappa^c_{ij}$ (vertical arrows) and acts as a template to be replicated
    (circular arrows) or transcribed (horizontal arrows) at an identical rate
    $w^c_{ij}$. This rate increases with an increase in the average catalytic
    activity molecules in protocell $i$ (denoted as
    $\langle\kappa^c_{i\tilde{j}}\rangle$ in the main text), but decreases with
    an increase in the molecule's own catalytic activity $\kappa^c_{ij}$,
    reflecting the trade-off between catalysing and templating functions.}
\end{figure}

The third assumption, that multilevel selection operates, posits that the
population of molecules is compartmentalised into primordial cells, or
protocells for short (Fig.\,\ref{fig:model}a), and that natural selection
operates at two levels:
\begin{itemize}
    \item within-cell selection, which arises from variations in replication
          rates across different molecules within the same cell;
    \item between-cell selection, which arises from variations in the average
          replication rates of molecules across different cells.
\end{itemize}
Multilevel selection is necessary because the provision of catalysis, which is
altruistic, cannot evolve in a well-mixed population, where each molecule
interacts equally likely with all others \citep{Takeuchi2012}.

A simple mathematical model incorporating the above assumptions can be
formulated as follows \citep{Takeuchi2019}:
\begin{itemize}
    \item Molecules exist in one of two stages, denoted as $\text{P}$ and
          $\text{Q}$ (Fig.\,\ref{fig:model}b).
    \item A molecule $j$ within protocell $i$ has a catalytic activity
          $\kappa^c_{ij}$ when in stage $c$ (Fig.\,\ref{fig:model}b).
    \item The rate at which the molecule acts as a template to produce a new
          molecule decreases with an increase in its catalytic activity
          $\kappa^c_{ij}$, reflecting the trade-off between catalysing and
          templating functions.
    \item This production rate increases with an increase in the average
          catalytic activity of molecules in protocell $i$, denoted as
          $\langle\kappa^\text{P}_{i\tilde{j}}\rangle$ and
          $\langle\kappa^\text{Q}_{i\tilde{j}}\rangle$, where the tildes in the
          angular brackets mark the indices over which the average is taken
          (this notational convention also applies to other symbols).
    \item A newly produced molecule either remains in the same stage as the
          template molecule (considered ``replication'') or transition to the
          other stage (considered ``transcription''). For simplicity,
          replication and transcription occur at the same rate, denoted
          $\omega^c_{ij}$ (Fig.\,\ref{fig:model}b). No other types of stage
          transition occur.
    \item For simplicity, the mutation of molecules and the growth and division
          of protocells are not explicitly modelled. Instead, it is assumed that
          variances in catalytic activity $\kappa^c_{ij}$ within protocells
          (i.e., the variance of $\kappa^c_{ij}$ across $j$ within $i$, averaged
          over all $i$) and between protocells (i.e., the variance of
          $\langle\kappa^c_{i\tilde{j}}\rangle$ across $i$) are maintained at
          $\sigma^2_\text{w}$ and $\sigma^2_\text{b}$, respectively.
\end{itemize}

The fitness of a molecule (denoted $\lambda_{ij}$) is defined as the growth rate
of a hypothetical population generated through the replication and transcription
of this molecule (i.e., its descendants) without mutation. To calculate
$\lambda_{ij}$, let $n^c_{ij}(\tau)$ be the number of molecules in stage $c$ in
this hypothetical population at time $\tau$. Since the number of replication or
transcription events per unit time is $w^c_{ij}$, the following equation
describes the growth of this population:
\begin{equation}
    \label{EqPopulationMatrix}
    \begin{bmatrix}
        n^\text{P}_{ij}(\tau+1) \\
        n^\text{Q}_{ij}(\tau+1)
    \end{bmatrix} =
    \begin{bmatrix}
        \omega^\text{P}_{ij} & \omega^\text{Q}_{ij} \\
        \omega^\text{P}_{ij} & \omega^\text{Q}_{ij}
    \end{bmatrix}
    \begin{bmatrix}
        n^\text{P}_{ij}(\tau) \\
        n^\text{Q}_{ij}(\tau)
    \end{bmatrix}
\end{equation}
The fitness $\lambda_{ij}$ corresponds to the largest eigenvalue of the $2
    \times 2$ matrix on the right-hand side of Eq.\,(\ref{EqPopulationMatrix}),
    which is calculated as:
\begin{equation}
    \label{eq:eigenvalue}
    \lambda_{ij} = \omega^\text{P}_{ij} + \omega^\text{Q}_{ij}.
\end{equation}

\subsection{Analysis of the model}
While the model incorporates the possibility of catalytic asymmetry as a
disparity between $\kappa^\text{P}_{ij}$ and $\kappa^\text{Q}_{ij}$, it also
accounts for transmission asymmetry, represented by a disparity between
$\omega^\text{P}_{ij}$ and $\omega^\text{Q}_{ij}$. To show this, we consider the
population of molecules described by Eq.\,(\ref{EqPopulationMatrix}). Let
$u^c_{ij}$ be the number of molecules (including both P and Q) generated through
the replication or transcription of molecule $j$ in stage $c$ within protocell
$i$ over a time period $T$. As $T \to \infty$, the normalised value of
$u^c_{ij}$ converges to a quantity known as Fisher's reproductive value, which
quantifies the stage $c$'s relative contribution to the future population
\citep{Rice2004}. If a molecule in stage $c$ is responsible for the perpetuation
of information---that is, it performs genic function---then $u^c_{ij} > 0$. If
it does not, then $u^c_{ij} = 0$. Therefore, the evolution of transmission
asymmetry corresponds to the evolution of reproductive value asymmetry
\citep{Takeuchi2019}. Mathematically, these reproductive values are the elements
of the left eigenvector associated with the largest eigenvalue $\lambda_{ij}$ of
the matrix on the right-hand side of Eq.\,(\ref{EqPopulationMatrix})
\citep{Rice2004}. Thus, it can be shown that
\begin{equation}
    \label{eq:left-eigenvector}
    \begin{split}
        u^\text{P}_{ij} & = \omega^\text{P}_{ij}, \\
        u^\text{Q}_{ij} & = \omega^\text{Q}_{ij}.
    \end{split}
\end{equation}
Equation~(\ref{eq:left-eigenvector}) aligns with the intuitive idea that the
more frequently a molecule replicates, the greater its contribution to
information transmission (however, the specific form of
Eq.\,(\ref{eq:left-eigenvector}) relies on the simplifying assumption that the
rates of transcription and replication are equal).

To summarise, catalytic asymmetry is modelled as a disparity between
$\kappa^\text{P}_{ij}$ and $\kappa^\text{Q}_{ij}$, while transmission asymmetry
is modelled as that between $\omega^\text{P}_{ij}$ and $\omega^\text{Q}_{ij}$.
These asymmetries are interdependent because an increase in $\kappa^c_{ij}$
reduces $\omega^c_{ij}$ due to the trade-off between catalytic and template
functions. In other words, catalytic asymmetry gives rise to transmission
asymmetry.

To understand how catalytic activity evolves, we calculate the change in the
average catalytic activity of the system per unit time, denoted
$\Delta\langle\kappa^c_{\tilde{i}\tilde{j}}\rangle$. Since each molecule leaves
the number of descendants proportional to its fitness per unit time, the
following equation holds:
\begin{equation}
    \Delta\langle\kappa^c_{\tilde{i}\tilde{j}}\rangle = \frac{\sum_{ij} \lambda_{ij} \kappa^c_{ij}}{\sum_{ij} \lambda_{ij}} - \langle\kappa^c_{\tilde{i}\tilde{j}}\rangle.
\end{equation}
Under the assumptions that the between-cell variance $\sigma^2_\text{b}$ and the
within-cell variance $\sigma^2_\text{w}$ of catalytic activity are sufficiently
small and that $\kappa^\text{P}_{ij}$ and $\kappa^\text{Q}_{ij}$ are
uncorrelated as $i$ and $j$ are varied,
$\Delta\langle\kappa^c_{\tilde{i}\tilde{j}}\rangle$ can be approximated as
\begin{subequations}
    \label{EqPriceGeneralFnckinsym}
    \begin{align}
        \label{EqPriceGeneralFnckinsym_P}
        \Delta\langle\kappa^\text{P}_{\tilde{i}\tilde{j}}\rangle & \approx\frac{\langle\omega^\text{P}_{\tilde{i}\tilde{j}}\rangle\left(\beta\sigma^2_\text{b}-\gamma\sigma^2_\text{w}\right)+\langle\omega^\text{Q}_{\tilde{i}\tilde{j}}\rangle\beta\sigma^2_\text{b}}{\langle\lambda_{\tilde{i}\tilde{j}}\rangle}, \\
        \label{EqPriceGeneralFnckinsym_Q}
        \Delta\langle\kappa^\text{Q}_{\tilde{i}\tilde{j}}\rangle & \approx\frac{\langle\omega^\text{P}_{\tilde{i}\tilde{j}}\rangle\beta\sigma^2_\text{b}+\langle\omega^\text{Q}_{\tilde{i}\tilde{j}}\rangle\left(\beta\sigma^2_\text{b}-\gamma\sigma^2_\text{w}\right)}{\langle\lambda_{\tilde{i}\tilde{j}}\rangle}.
    \end{align}
\end{subequations}
These approximations hold to the linear terms of the second central moments of
$\kappa^c_{ij}$, ignoring the non-linear terms of these moments and the
higher-order moments (see \citet{Takeuchi2019} for details):

Equations~(\ref{EqPriceGeneralFnckinsym}) have the following meaning. Symbols
$\beta$ and $\gamma$ represent selection pressures at different levels
($\beta\geq0$ and $\gamma\geq0$). Specifically, $\beta$ is the strength of
between-cell selection acting on catalytic activity. The positiveness of the
terms involving $\beta$ means that between-cell selection favours an increase in
catalytic activity. By contrast, $\gamma$ is the strength of within-cell
selection acting on catalytic activity. The negativeness of the terms involving
$\gamma$ means that within-cell selection favours a decrease in catalytic
activity. This negative effect comes from the fact that increasing the catalytic
activity of a molecule reduces this molecule's relative replication
rate within a protocell, as providing catalysis is altruistic. Together, these
opposing forces cause conflicting multilevel selection, a process in which
selection operates antagonistically at two levels.

The terms involving $\langle\omega^\text{P}_{\tilde{i}\tilde{j}}\rangle$ and
$\langle\omega^\text{Q}_{\tilde{i}\tilde{j}}\rangle$ in
Eqs.\,(\ref{EqPriceGeneralFnckinsym}) represent different transmission pathways
through which evolution occurs. Specifically, the terms involving
$\langle\omega^\text{P}_{\tilde{i}\tilde{j}}\rangle$ represent contributions to
evolution through the descendants of P molecules, i.e., through P's information
transmission. By contrast, the terms containing
$\langle\omega^\text{Q}_{\tilde{i}\tilde{j}}\rangle$ represent contributions to
evolution through the descendants of Q molecules, i.e., through Q's information
transmission.

Equations~(\ref{EqPriceGeneralFnckinsym}) imply a positive feedback loop between
the evolution of catalytic asymmetry and that of transmission asymmetry,
suggesting spontaneous symmetry breaking. To illustrate this, we begin with a
system that is symmetric with respect to both catalysis and transmission:
$\langle\kappa^\text{P}_{\tilde{i}\tilde{j}}\rangle =
\langle\kappa^\text{Q}_{\tilde{i}\tilde{j}}\rangle$ and
$\langle\omega^\text{P}_{\tilde{i}\tilde{j}}\rangle =
\langle\omega^\text{Q}_{\tilde{i}\tilde{j}}\rangle$. In this state, P and Q
molecules are functionally identical, so that changes in their catalytic
activities are identical too:
$\Delta\langle\kappa^\text{P}_{\tilde{i}\tilde{j}}\rangle =
\Delta\langle\kappa^\text{Q}_{\tilde{i}\tilde{j}}\rangle$. Next, we introduce a
slight perturbation to catalytic symmetry, due to mutations or random genetic
drift, such that P's catalytic activity is marginally increased compared to Q:
\begin{equation}
    \label{eq:catalytic-asymmetry}
    \langle\kappa^\text{P}_{\tilde{i}\tilde{j}}\rangle > \langle\kappa^\text{Q}_{\tilde{i}\tilde{j}}\rangle.
\end{equation}
This increase comes at a cost to P's templating activity, reducing its
transmission compared to Q:
\begin{equation}
    \label{eq:transmission-asymmetry}
    \langle\omega^\text{P}_{\tilde{i}\tilde{j}}\rangle < \langle\omega^\text{Q}_{\tilde{i}\tilde{j}}\rangle.
\end{equation}
The resulting transmission asymmetry further amplifies catalytic asymmetry
\begin{equation}
    \label{eq:spontaneous-symmetry-breaking}
    \Delta\langle\kappa^\text{P}_{\tilde{i}\tilde{j}}\rangle > \Delta\langle\kappa^\text{Q}_{\tilde{i}\tilde{j}}\rangle,
\end{equation}
as explained in the next paragraph.

Equation~(\ref{eq:spontaneous-symmetry-breaking}) follows from the fact that
Eqs.\,(\ref{EqPriceGeneralFnckinsym}) involve asymmetric coupling between
conflicting multilevel selection (i.e., the terms involving $\beta$ and
$\gamma$) and different transmission pathways (i.e., the terms involving
$\langle\omega^\text{P}_{\tilde{i}\tilde{j}}\rangle$ and
$\langle\omega^\text{Q}_{\tilde{i}\tilde{j}}\rangle$). Specifically, in
Eq.\,(\ref{EqPriceGeneralFnckinsym_P}), $\gamma$ appears in the term containing
$\langle\omega^\text{P}_{\tilde{i}\tilde{j}}\rangle$ but not in the term
containing $\langle\omega^\text{Q}_{\tilde{i}\tilde{j}}\rangle$. This means that
within-cell selection tends to decrease P's catalytic activity through P's
descendants but not through Q's descendants. This is because an increase in the
catalytic activity of a P molecule reduces the replication rate of this molecule
but not the replication rate of its transcript, Q. Likewise, in
Eq.\,(\ref{EqPriceGeneralFnckinsym_Q}), $\gamma$ appears in the term containing
$\langle\omega^\text{Q}_{\tilde{i}\tilde{j}}\rangle$ but not in the term
containing $\langle\omega^\text{P}_{\tilde{i}\tilde{j}}\rangle$, permitting a
similar interpretation as above. In contrast, $\beta$ appears in both terms
containing $\langle\omega^\text{P}_{\tilde{i}\tilde{j}}\rangle$ and terms
containing $\langle\omega^\text{P}_{\tilde{i}\tilde{j}}\rangle$ in
Eqs.\,(\ref{EqPriceGeneralFnckinsym_P}) and (\ref{EqPriceGeneralFnckinsym_Q}).
This is because an increase in the average catalytic activity of molecules
(whether P or Q) in a protocell increases the replication rate of all molecules
in that protocell. This asymmetric coupling between multilevel selection and
transmission pathways means that when P is replicated less frequently than Q
(i.e., $\langle\omega^\text{P}_{\tilde{i}\tilde{j}}\rangle <
\langle\omega^\text{Q}_{\tilde{i}\tilde{j}}\rangle$), within-cell selection on
P's catalytic activity through P's descendants (i.e.,
$\langle\omega^\text{P}_{\tilde{i}\tilde{j}}\rangle\gamma\sigma^2_\text{w}$) has
less impact than within-cell selection on Q's catalytic activity through Q's
descendants (i.e.,
$\langle\omega^\text{Q}_{\tilde{i}\tilde{j}}\rangle\gamma\sigma^2_\text{w}$).
Therefore, Eq.\,(\ref{eq:spontaneous-symmetry-breaking}) follows. The result
that Eq.\,(\ref{eq:spontaneous-symmetry-breaking}) follows from
Eq.\,(\ref{eq:catalytic-asymmetry}) via Eq.\,(\ref{eq:transmission-asymmetry})
indicates a positive feedback loop, whereby catalytic asymmetry and transmission
asymmetry reinforce each other.

Whether the feedback loop described in the previous paragraph induces symmetry
breaking between P and Q depends on the relative magnitudes of within-cell and
between-cell variances, $\sigma^2_\text{w}$ and $\sigma^2_\text{b}$,
respectively. Specifically, under the assumption that within-cell and
between-cell selection pressures are of similar magnitudes ($\beta\sim\sigma$),
if $\sigma^2_\text{w}\ll\sigma^2_\text{b}$,
Eqs.\,(\ref{EqPriceGeneralFnckinsym}) simplify to
\begin{subequations}
    \begin{align}
        \Delta\langle\kappa^\text{P}_{\tilde{i}\tilde{j}}\rangle & \approx\beta\sigma^2_\text{b}, \\
        \Delta\langle\kappa^\text{Q}_{\tilde{i}\tilde{j}}\rangle & \approx\beta\sigma^2_\text{b},
    \end{align}
\end{subequations}
up to the linear terms of the second central moments of $\kappa^c_{ij}$. These
equations indicate that between-cell selection dominates evolutionary dynamics,
preventing spontaneous symmetry breaking.

To examine whether spontaneous symmetry breaking occurs when $\sigma^2_\text{w}$
is sufficiently large relative to $\sigma^2_\text{b}$, we consider the following
specific definition of $\omega^c_{ij}$:
\begin{align}
    \label{EqOmegaInfkinsym}
    \omega^c_{ij}
    =e^{\langle\kappa^\text{P}_{i\tilde{j}}\rangle +
    \langle\kappa^\text{Q}_{i\tilde{j}}\rangle} \frac{e^{-\kappa^c_{ij}}}{\langle
    e^{-\kappa^\text{P}_{i\tilde{j}}}\rangle+\langle e^{-\kappa^\text{Q}_{i\tilde{j}}}\rangle},
\end{align}
where $c$ is P or Q \citep{Takeuchi2019}. This definition, which was arbitrarily
chosen to balance simplicity and the need to satisfy the model assumptions
listed earlier, involves three factors. The first factor
$e^{\langle\kappa^\text{P}_{i\tilde{j}}\rangle+\langle\kappa^\text{Q}_{i\tilde{j}}\rangle}$
means that $\omega^c_{ij}$ increases with an increase in the average catalytic
activity in protocell $i$. The second factor $e^{-\kappa^c_{ij}}$ means that
$\omega^c_{ij}$ decreases with an increase in the molecule's catalytic activity,
representing the cost of providing catalysis. The last factor normalises this
cost so that the per-cell average fitness of molecules
$\langle\lambda_{i\tilde{j}}\rangle$ has a simple expression, namely,
$e^{\langle\kappa^\text{P}_{i\tilde{j}}\rangle +
\langle\kappa^\text{Q}_{i\tilde{j}}\rangle}$.
    
\begin{figure}[tb!]
    \centering
    \includegraphics{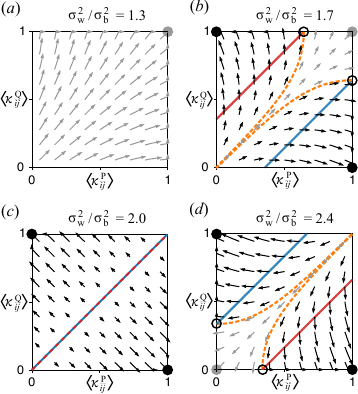}
    \caption{\label{fig:phase-portrait}Phase plane analyses of
    Eqs.\,(\ref{EqForPhasePortrait}) for different ratios of $\sigma^2_\text{w}$
    and $\sigma^2_\text{b}$. To perform these analyses,
    Eqs.\,(\ref{EqForPhasePortrait}) were adapted by replacing $\Delta$ with
    time derivative $\frac{\text{d}}{\text{d}\tau}$ and setting
    $\frac{\text{d}}{\text{d}\tau}\langle\kappa^c_{\tilde{i}\tilde{j}}\rangle=0$
    when $\langle\kappa^c_{\tilde{i}\tilde{j}}\rangle=0$ or $1$ so that
    $\langle\kappa^c_{\tilde{i}\tilde{j}}\rangle$ is bounded. The solid lines
    represent the nullclines of
    $\langle\kappa^\text{P}_{\tilde{i}\tilde{j}}\rangle$ (red) and
    $\langle\kappa^\text{Q}_{\tilde{i}\tilde{j}}\rangle$ (blue). The filled
    circles indicate symmetric (grey) and asymmetric (black) stable equilibria;
    the open circles, unstable equilibria. The arrows indicate vector fields
    leading to symmetric (grey) or asymmetric (black) equilibria. The dashed
    lines (orange) demarcate the basins of attraction. \textbf{(a)} Within-cell
    variance is so small that between-cell selection completely dominates,
    maximising both $\langle\kappa^\text{P}_{\tilde{i}\tilde{j}}\rangle$ and
    $\langle\kappa^\text{Q}_{\tilde{i}\tilde{j}}\rangle$. \textbf{(b)}
    Within-cell variance is large enough to create stable asymmetric equilibria.
    However, the equilibrium maximising both
    $\langle\kappa^\text{P}_{\tilde{i}\tilde{j}}\rangle$ and
    $\langle\kappa^\text{Q}_{\tilde{i}\tilde{j}}\rangle$ is still present and
    stable. Spontaneous symmetry breaking occurs when the initial values of
    $\langle\kappa^\text{P}_{\tilde{i}\tilde{j}}\rangle$ and
    $\langle\kappa^\text{Q}_{\tilde{i}\tilde{j}}\rangle$ are sufficiently low.
    \textbf{(c)} The critical point, where the nullclines overlap. \textbf{(d)}
    Within-cell variance is so large that the equilibrium maximising both
    $\langle\kappa^\text{P}_{\tilde{i}\tilde{j}}\rangle$ and
    $\langle\kappa^\text{Q}_{\tilde{i}\tilde{j}}\rangle$ does not exist.
    Spontaneous symmetry breaking occurs when the initial values of
    $\langle\kappa^\text{P}_{\tilde{i}\tilde{j}}\rangle$ and
    $\langle\kappa^\text{Q}_{\tilde{i}\tilde{j}}\rangle$ are sufficiently high. The figure is reproduced from \citet{Takeuchi2019}}
\end{figure}

Under the definition in Eq.\,(\ref{EqOmegaInfkinsym}),
Eqs.\,(\ref{EqPriceGeneralFnckinsym}) can be transformed as
\begin{subequations}
    \label{EqForPhasePortrait}
    \begin{align}
        \Delta\langle\kappa^\text{P}_{\tilde{i}\tilde{j}}\rangle \approx
        \sigma^2_\text{b}-\sigma^2_\text{w}\frac{\langle e^{-\kappa^\text{P}_{i\tilde{j}}}\rangle}{\langle
        e^{-\kappa^\text{P}_{i\tilde{j}}}\rangle+\langle e^{-\kappa^\text{Q}_{i\tilde{j}}}\rangle}, \\
        \Delta\langle\kappa^\text{Q}_{\tilde{i}\tilde{j}}\rangle\approx
        \sigma^2_\text{b}-\sigma^2_\text{w}\frac{\langle e^{-\kappa^\text{Q}_{i\tilde{j}}}\rangle}{\langle
        e^{-\kappa^\text{P}_{i\tilde{j}}}\rangle+\langle e^{-\kappa^\text{Q}_{i\tilde{j}}}\rangle},
    \end{align}
\end{subequations}
again up to the linear terms of the second central moments of $\kappa^c_{ij}$
\citep{Takeuchi2019}. Equations~\ref{EqForPhasePortrait} were investigated
through phase-plane analyses for different ratios of $\sigma^2_\text{w}$ and
$\sigma^2_\text{b}$, as displayed in Fig.\,\ref{fig:phase-portrait}
\citep{Takeuchi2019}. These analyses reveal that when $\sigma^2_\text{w}$ is
sufficiently large relative to $\sigma^2_\text{b}$, P and Q can undergo
spontaneous symmetry breaking, differentiating into molecules that provide
catalysis but are not replicated or transcribed (``enzymes'') and those that are
replicated and transcribed but do not provide catalysis (``genes''). This result
suggests that the distinction between genes and enzymes can evolve as a
consequence of multilevel selection operating on a stage-structured population
of replicating catalytic molecules, provided $\sigma^2_\text{w}$ is sufficiently
large relative to $\sigma^2_\text{b}$.

The condition that $\sigma^2_\text{w}$ is sufficiently large relative to
$\sigma^2_\text{b}$, by definition, means that the variance of catalytic
activity between molecules within protocells is sufficiently large relative to
the variance of catalytic activity between protocells. This condition is
fulfilled when the number of molecules per protocell is large, the mutation rate
of molecules is high, or both \citep{Takeuchi2019,Takeuchi2022}. This condition
corresponds to low genetic relatedness between molecules---specifically, this
relatedness is equal to
$\sigma^2_\text{b}/(\sigma^2_\text{w}+\sigma^2_\text{b})$
\citep{Takeuchi2019,Takeuchi2022}. Therefore, the model reviewed above predicts
that a gene-enzyme distinction evolves when genetic relatedness is low.

The mathematical results described above agree with the
computational simulations of an individual-based model that explicitly
incorporates chemical reaction, mutation, resource limitation, and
compartmentalisation of molecules \citep{Takeuchi2019}.

The model reviewed above differs from previous models in two key respects. First, our model explicitly demonstrates the differentiation of molecules into those that transmit information (genes) and those that provide catalysis (enzymes) through spontaneous symmetry breaking. Some of the previous models do not allow any symmetry breaking \citep{Frank1994,Levin2017}. Other models demonstrate catalytic symmetry breaking but do not demonstrate the unidirectional flow of information between catalysts and non-catalysts \citep{Takeuchi2008a,Takeuchi2017}, since they assume that catalytic asymmetry occurs between complementary strands of RNA molecules \citep{Hogeweg2003,Boza2014}. Another model demonstrates the evolution of this unidirectional information flow but does not demonstrate catalytic symmetry breaking, since it predefines one type of molecules as a catalyst and the other as a non-catalyst \citep{Takeuchi2011}.

The second key difference with our model concerns a condition for the evolution of gene-enzyme division. Our model predicts that this evolution occurs when relatedness is sufficiently low, whereas the model of \citet{Michod1983} predicts that it occurs when relatedness is sufficiently high. This contrast reflects a fundamental difference in the underlying mechanisms for this evolution. Our model relies on the positive feedback loop emerging from conflicting multilevel selection acting on a stage-structured population. In contrast, Michod's model relies on competition between two species of replicators---catalyst (altruist) and non-catalyst (cheater)---in a subdivided population. Michod's model predicts that the catalyst is favoured when relatedness is sufficiently high \citep{Michod1983}. This outcome is interpreted as indicating that the division of labour between genes and enzymes is favoured when relatedness is sufficiently high \citep{Michod1983}, although this division is not explicitly included in the model.

Finally, we add that the division of labour between genes and enzymes has been demonstrated to evolve through spontaneous symmetry breaking even when molecules are not compartmentalised into protocells but spatially distributed on a surface \citep{Fu2024}. In this case, the division of labour evolves when the diffusion constant of molecules is sufficiently high, a condition that corresponds to low relatedness. This means that the division of labour between genes and enzymes does not necessarily depend on the origin of protocells.

\section{Evolutionary consequences of the Central Dogma}
The emergence of the division of labour between information transmission and
expression can enhance the evolvability of protocells in three ways. First, this
division likely induces selection to increase the copy number of enzymes per
protocell relative to genes---i.e., numerical asymmetry---because increasing
enzyme abundance enhances the productivity of protocells. This trend has been
observed in computer simulations \citep{Takeuchi2019}. The resulting decrease in
the relative copy number of genes amplifies the effects of mutations on the
protocell's phenotype, thereby increasing its evolvability \citep{Kaneko2002}.
Specifically, a low gene copy number allows mutations to cause large phenotypic
variation, whereas a high gene copy number averages out the effects of mutations
due to the law of large number \citep{Koch1984, Kaneko2002}. Notably, similar
numerical asymmetry is observed at the other levels of the biological hierarchy,
such as small numbers of germline cells per multicellular organism and queens
per eusocial colony.

Second, in the context of number symmetry breaking in the generalized central dogma, the replication of minority elements (e.g., DNA) occurs at a significantly slower rate than that of majority elements (e.g., proteins) \citep{Kaneko2002}. More broadly, evolving dynamical systems often exhibit a trend where elements with slower time scales regulate the behavior of those with faster dynamics. This separation of time scales also facilitates evolution \citep{Kohsokabe2016,Furusawa2018}. Slower elements typically function as information carriers (genotype), while faster elements act as expressors (phenotype) \citep{Kaneko2024}, a principle also explored in the context of multilevel learning \citep{Vanchurin2022a}.

Third, the division of labour between information transmission and expression
enhances the evolvability of protocells by enabling the regulation of
information expression. Without this division, identical molecules both transmit
and express information, forcing information to be expressed whenever it is
transmitted. By separating these functions, the division of labour enables
control over information expression. This ability likely enhances the cell's
capacity to adapt to changing environments.

\section{Open questions}
Finally, we highlight several open questions that arise from reconceptualising
the Central Dogma as a division of labour between information transmission and
expression.

First, DNA and protein molecules are linked through genetic code in reality. How
does such a symbolic relationship emerge?

Second, the division of labour between information transmission and expression
evolved independently across vastly different biological scales. Can this
recurrent pattern be explained by the same symmetry-breaking mechanism described
above? How does this mechanism differ from those proposed for the evolution of the germline-soma distinction and eusociality \citep{Michod2006,Barton2007,Johnstone2008,Davies2012c,Rueffler2012,Cooper2018}?

Third, the Central Dogma in biological systems is manifested by DNA and protein,
with RNA serving as an intermediary. How do the specific chemical properties and
roles of these materials influence or interact with the proposed
symmetry-breaking mechanism?

\bibliographystyle{abbrvnat}
\bibliography{jabref.bib}
\end{document}